\documentclass[twocolumn,showpacs,prl,superscriptaddress]{revtex4}

\usepackage{amsmath}
\usepackage{bm}
\usepackage{array}
\usepackage{graphicx}
\usepackage{dcolumn}   
\usepackage{amssymb}
\usepackage[english]{babel}

\begin{document}

\preprint{PRL}
\title{Vibrations of a Columnar Vortex in a Trapped Bose-Einstein Condensate}
\date{\today}
\author{Lyndon Koens}
\affiliation{School of Physics, The University of Melbourne, Parkville,
3010, Australia}
\author{Tapio P. Simula}
\affiliation{School of Physics, Monash University, Clayton,
3800, Australia}
\author{Andrew M. Martin}
\affiliation{School of Physics, The University of Melbourne, Parkville,
3010, Australia}
\begin{abstract}
We derive a governing equation for a Kelvin wave supported on a vortex line in a Bose-Einstein condensate, in a rotating cylindrically symmetric parabolic trap. From this solution the Kelvin wave dispersion relation is determined. In the limit of an oblate trap and in the absence of longitudinal trapping our results are consistent with previous work. We show that the derived Kelvin wave dispersion in the general case is in quantitative agreement with numerical calculations of the Bogoliubov spectrum and offer a significant improvement upon previous analytical work.     
\end{abstract}
\maketitle
The behavior of turbulent flows, tornadoes, mixing processes, synoptic scale weather phenomena and sunspots all critically depend on our understanding of vortex dynamics. Quantitative investigation of vortices started in the mid 1800's with the development of the Navier-Stokes equation, with the properties of a vortex being described by streamlines of vorticity.  In 1880, Thomson (Lord Kelvin) determined the dispersion relation  \cite{Philos.Mag.10.155.} for  a specific excitation on a vortex line. This excitation takes the form of a chiral normal mode in which the perturbation propagates along a vortex line and rotates about its unperturbed position, distorting the vortex line into a helical shape.  



In superfluids these vortices differ from their classical counterparts by having quantized circulation and a single line of vorticity (a vortex line) associated with them \cite{donnelly2005quantized}. A quantised vortex line, in analogy with a classical vortex, supports Kelvin waves \cite{donnelly2005quantized,Proc.R.Soc.245.546,Sov.Phys.JETP.10.227,Sov.Phys.JETP.13.451,Phys.Rev.Lett.42.1062,Rev.Mod.Phys.59.87,Astrophys.J.387.276,PhysRevLett.90.100403,Phys.Rev.A.69.043617,Rev.Mod.Phys.81.647}. The first experimental investigations of Kelvin waves, in  a superfluid, were in carried out in cryogenically cooled helium \cite{donnelly2005quantized,Proc.R.Soc.245.546,Sov.Phys.JETP.10.227,Phys.Rev.Lett.42.1062}. More recently Bose-Einstein condensates (BECs) have provided a new platform to  investigate the properties of quantized vortices \cite{PhysRevLett.83.2498,PhysRevLett.85.2857,PhysRevLett.84.806, PhysRevLett.85.2223,Science.292.476,Phys.Rev.Lett.88.010405,J.Low.Temp.Phys.161.574}. The highly controllable nature of BEC systems has enabled the experimental investigation of Kelvin waves on  a single vortex line \cite{PhysRevLett.90.100403}. The behaviour of such waves has been investigated \cite{PhysRevLett.90.180401, PhysRevLett.101.020402, PhysRevA.78.053604} and plays a crucial role in understanding the details of superfluid turbulence \cite{Phys.Rev.Lett.92.035301,JLowTempPhys.145.7,ProgLowTempPhysQT}. In trapped systems the Kelvin wave dispersion for a single vortex line has been obtained numerically, via solving the Bogoliubov spectrum for a single vortex line \cite{J.Phys.Soc.Jpn.66.3502,PhysRevLett.101.020402, PhysRevA.78.053604}. The Kelvin wave dispersion relation in the limit of long wavelengths and in the absence of trapping is \cite{Sov.Phys.JETP.13.451}
\begin{equation} \label{hd}
\omega = \frac{ \hbar k^{2}}{2 M} \ln\left(\frac{1}{|k| r_{c}}\right),
\end{equation}
where $\omega$ is the excitation frequency of the mode, $k$ is the wavenumber, $M$ is the particle mass and $r_{c}$ is the vortex core parameter. 
 

A general formulation for a quantised vortex line in a trapped rotating BEC has proven difficult. Most methods have relied on matched asymptotic expansions \cite{PhysicaD78.1994.1, PhysicaD.47.1991.353, PhysRevA.86.013605,Phys.Rev.A.62.063617}. Koens and Martin used such a procedure to obtain a set of equations that describe the behaviour of a perturbed vortex line \cite{PhysRevA.86.013605}. In this analysis the general equations [Eqs. (68) and (69) in Ref.~\cite{PhysRevA.86.013605}] contained an undetermined {\it constant}. Here we eliminate the constant to obtain a single equation that describes the radial position of a vortex line, supporting a Kelvin wave. We then determine the general solutions of this equation, enabling us to make favorable comparisons with previous results in limiting regimes \cite{PhysRevLett.84.5919, PhysRevA.63.043608,PhysRevA.64.043611,PhysRevA.64.053611, PhysRevA.86.013605}. From the general solution the Kelvin wave dispersion relation, for a single vortex line in a BEC in a cylindrically symmetric parabolic trap, is determined. This dispersion  quantitatively agrees with numerically calculated Bogoliubov spectra \cite{PhysRevA.78.053604},  in contrast to  previous analytic  results \cite{Phys.Rev.A.62.063617}.

As shown in Ref.~\cite{PhysRevA.86.013605,NoteLyndon} the equations governing the positional dependence of the vortex line can be obtained. In cylindrical coordinates, defined by the radial, $\hat{\rho}$, angular, $\hat{\phi}$, and axial, $\hat{z}$, unit vectors, these equations are:
\begin{widetext}
\begin{align}
\frac{\partial \rho}{\partial t}  &=  - \frac{\hbar \rho \left[\ln( r_{c})+1\right] }{2  M g |\Psi_{TF}|^{2} }  \frac{\partial \phi}{\partial z} \frac{\partial V_{tr}}{\partial z}  
-\frac{\hbar \left[\ln \left( \frac{1}{r_{c}} \sqrt{\frac{ \rho}{\rho \left(\partial_{z} \phi\right)^{2} - \partial_{z}^{2} \rho} }\right)+1\right] }{2 M} \left( \rho \frac{\partial^{2} \phi}{\partial z^{2}} + 2 \frac{\partial \rho}{\partial z}  \frac{\partial \phi}{\partial z} \right) - \frac{3\hbar  \left[\ln\left( \frac{R_{\perp}}{ r_{c}} \right)+\frac{2}{3}\right] \mbox{ }  }{4 \rho M g |\Psi_{TF}|^{2} } \frac{\partial V_{tr}}{\partial \phi} \nonumber \\ 
&+ 2 \frac{\Omega }{\rho \nabla_{\perp}^{2} V_{tr}} \frac{\partial V_{tr}}{\partial \phi} + \hat{\rho}\cdot \mathbf{E}\left( \frac{ - \rho \frac{\partial \phi}{\partial z} \hat{\rho} + \frac{\partial \rho}{\partial z} \hat{\phi} }{g |\Psi_{TF}|^{2}} \frac{\partial V_{tr}}{\partial z}, \mathbf{r} \right),  \label{rho}
\end{align}
\begin{align}
\rho \frac{\partial\phi}{\partial t}  &=  \quad \frac{\hbar \ln( r_{c})}{2 M g |\Psi_{TF}|^{2} }  \frac{\partial \rho}{\partial z} \frac{\partial V_{tr}}{\partial z} + \frac{3\hbar  \ln\left( \frac{R_{\perp}}{ r_{c}} \right)  }{4 M g |\Psi_{TF}|^{2} } \frac{\partial V_{tr}}{\partial \rho} 
+\frac{\hbar \ln \left( \frac{1}{r_{c}} \sqrt{\frac{ \rho}{\rho \left(\partial_{z} \phi\right)^{2} - \partial_{z}^{2} \rho} }\right)}{2 M}  \left[\frac{\partial^{2} \rho}{\partial z^{2}} - \rho \left(\frac{\partial \phi}{\partial z}\right)^{2} \right]  \nonumber \\ 
&- 2 \frac{\Omega}{\nabla_{\perp}^{2} V_{tr}} \frac{\partial V_{tr}}{\partial \rho} + \hat{\phi}\cdot \mathbf{E}\left( \frac{ - \rho \frac{\partial \phi}{\partial z} \hat{\rho} + \frac{\partial \rho}{\partial z} \hat{\phi}}{g |\Psi_{TF}|^{2}} \frac{\partial V_{tr}}{\partial z}, \mathbf{r} \right), \label{phi}
\end{align}
\end{widetext}
where $\rho$ and $\phi$ describe the radial and angular coordinates of the vortex line at a given $z$ (the dependence on $z$ is implicit), $V_{tr}$ is the trapping potential, $|\Psi_{TF}|^{2}$ is the Thomas--Fermi condensate density, $g$ is the inter-particle interaction strength, $R_{\perp}$ is the radial Thomas--Fermi radius, $\Omega$ is the rotation frequency of the trap, and $\mathbf{E}$ is an unknown constant vector. In general, these equations relate the motion of the vortex line [left hand sides (LHS) of Eqs. \eqref{rho} and \eqref{phi}] to its distortion and coupling to the trapping potential [right hand sides (RHS) of Eqs. \eqref{rho} and \eqref{phi}]. The constant provides coupling between the motion of the vortex line and the trapping potential in the $z$-direction. For the equations to be dimensionally consistent, $\mathbf{E}$ needs to take a form similar to the first terms on the RHS of Eqs. \eqref{rho} and \eqref{phi}. Hence, each equation has the same constant, which can be eliminated.

In this work, we are interested in the properties of helical waves described by  $\phi = k z - \omega t$, in the presence of a harmonic trapping potential $V_{tr} (r_{\perp}=\rho, z)= M \omega_{\perp}^{2} \rho^{2} /2 + M \omega_{z}^{2} z^{2} /2$, where $\omega_{\perp}$ and $\omega_{z}$ are the trapping frequencies in the $r_{\perp}$ and $z$ directions, respectively. Applying these conditions with $\rho$ being time independent, the governing equation is
\begin{widetext}
\begin{eqnarray} \label{1eq}
\left[ \Omega - \omega  - \frac{3\hbar  \omega_{\perp}^{2} \ln\left( \frac{R_{\perp}}{ r_{c}} \right)  }{4  g |\Psi_{TF}|^{2}} +\frac{\hbar k^{2} \Theta}{M} \right] \rho^{2} 
 = - \frac{\hbar \omega_{z}^{2}  z \rho }{2 g |\Psi_{TF}|^{2}}  \frac{\partial \rho}{\partial z} + \frac{\hbar \rho\, \Theta}{M} \frac{\partial^{2} \rho}{\partial z^{2}} 
 - \frac{\hbar \left[2\Theta+1\right] }{ M}   \left(\frac{\partial \rho}{\partial z}\right)^{2},
\end{eqnarray}
\end{widetext}
where $\Theta=-\ln(r_c\sqrt{k^2-\partial_z^2\rho/\rho})/2$.
The LHS of Eq.~\eqref{1eq} defines the properties of a vortex line when $\rho$ is constant. The first of the terms on the RHS contains the influence from $z$ confinement and the other two represent the influence the curving of the vortex line in the radial direction has on its motion in $\phi$ and $\rho$  respectively.

Equation~\eqref{1eq} does not admit an analytic solution in a closed form. However, by assuming that: (i)  $\partial_{z}^{2} \rho /\rho$ is small compared to $k^{2}$  and may be approximated by a  constant $\beta$ and (ii) taking the effect of $g|\Psi_{TF}|^{2}$ to be a  constant, to leading order, defined by the condensate chemical potential, $\mu=g|\Psi_{TF}(0,0)|^{2}$, Eq.~\eqref{1eq} reduces to
\begin{equation} \label{s1eq}
\gamma \rho'^{2} = - z' \rho' \frac{\partial \rho'}{\partial z'} + \zeta \rho' \frac{\partial^{2} \rho'}{\partial z'^{2}} - \varepsilon \left(\frac{\partial \rho'}{\partial z'}\right)^{2},
\end{equation}
where $\zeta = -\ln(r_{c} \sqrt{k^{2} -\beta})/2$ 
and $\varepsilon  = 1+2 \zeta$.
In Eq.~\eqref{s1eq} $z' = z/R_{z} $, with $R_z$ being the axial Thomas--Fermi radius, $\rho' =\rho/ r_{c} $,
\begin{eqnarray}
\gamma =\frac{M}{\hbar} \Lambda^{2/5} \left[\Omega-\omega \right] + k^{2} \zeta \Lambda^{2/5} - \frac{3 }{ 10 \lambda^{2} } \ln \left(\frac{\Lambda}{r_{c}^{5}}\right),
  \end{eqnarray} 
  where $\Lambda= \frac{15 a N \hbar^{2}}{M^{2} \lambda^{4} \omega_{\perp}^{2} }$, with $a$ being the s-wave scattering length, $N$ the number of particles in the condensate and $ \lambda=\omega_z/ \omega_{\perp}$.


Making a change of variables such that $\rho'(z')$ has the form $ f(z')^{-\frac{\zeta}{\varepsilon -\zeta}}$ Eq.~\eqref{s1eq} reduces to 
\begin{equation}
\gamma (\zeta-\varepsilon ) f(z') + \zeta\left[ z' \frac{d f(z')}{d z'} -\zeta \frac{d^{2} f(z')}{d z'^{2}} \right] = 0,
\end{equation}
with the solution
\begin{equation} \label{gen}
\rho'(z') = \left[C_1 \mbox{ } H_{ A} \left( \frac{z'}{\sqrt{2 \zeta}}\right) + C_2 \mbox{ } _{1}F_{1}\left( -\frac{A}{2} ;\frac{1}{2};\frac{z'^{2}}{ 2 \zeta}\right)\right]^{-\frac{\zeta}{1+\zeta}},
\end{equation}
where $A = \gamma (\varepsilon -\zeta) /\zeta$, $H_{p}(z)$ is the Hermite polynomial of order $p$, $_{1}F_{1}(l;m;x)$ is the Kummer confluent hyper-geometric function, and $C_1$ and $C_2$ are integration constants.

Due to the axi-symmetry of the trapping potential, the vortex line structure described is physical when $\rho'$ is not multivalued within the condensate radius ($z = \pm R_{z}$), and is symmetric or anti-symmetric across $z=0$. Equation~\eqref{gen} can satisfy these conditions if $C_1 = 0$, reducing the solution to the symmetric hyper-geometric function or if $A$ is a positive even integer, making $H_{A}(x)$ symmetric. In general there is no physical reason to restrict $A$ to be an integer, hence we only consider the case where 
$C_1 = 0$. This solution gives $\rho'$ a ``U" shape, centered around $z'=0$, for example see Fig. 1(a). 
In this arrangement, an anti-symmetric vortex line has a higher energy than its symmetric counterpart, hence this solution does not support anti-symmetric structures. 

From Eq.~\eqref{gen} it is possible to check the assumption $|\beta|\ll k^2$. At $z=0$ $|\beta|$ is typically of order $10^{-9}$ $\mu$m$^{-2}$ for $k^{2} \approx 10^{-2}$ $\mu$m$^{-2}$. Nevertheless $\beta$ does depend on the ratio of the trapping frequencies, $\lambda$, with $\beta \rightarrow 0$ as $\lambda \rightarrow \infty $ and $\beta \rightarrow -\infty$ as $\lambda \rightarrow 0$. The divergence in $\beta$ as $\lambda \rightarrow 0$ is slow, indicating that $\beta$ plays a relatively insignificant role in Eq.~\eqref{1eq} unless the trap is extremely prolate. In general, we calculate $\beta$ self-consistently, through the definition $\beta=\partial^2_z \rho/\rho$.


{\it Oblate Limit --} Equation~\eqref{gen} simplifies  in the limits of extremely oblate ($\lambda \rightarrow \infty$) and prolate ($\lambda \rightarrow 0$) trapping potentials. In the oblate limit, $\gamma \rightarrow 0$, $\zeta \rightarrow - \ln\left(r_{c} |k|\right)/2$ and $\varepsilon \rightarrow 1-\ln\left(r_{c} |k|\right)$, 
Eq.~\eqref{s1eq} 
has the solution
\begin{equation} \label{linf}
\rho'(z') = \left[ C_1 \mbox{ }\left(1+\zeta\right) \sqrt{2 \pi}  \frac{\mbox{erf}\left(\frac{i z'}{\sqrt{2 \zeta}}\right)}{i} + 2 \sqrt{\zeta} \mbox{ } C_2 \right]^{-\frac{\zeta}{1+\zeta}},
\end{equation}
where $C_1$ and $C_2$ are again integration constants and $\mbox{erf}(x)$ is the error function. Applying the symmetry and divergence conditions $\rho'=(2 \sqrt{\zeta} \mbox{ } C_2)^{-\frac{\zeta}{1+\zeta}}$. If $\rho'$ is finite, from Eq.~\eqref{s1eq} the Kelvin wave has the frequency
\begin{equation} \label{dro}
\omega_{2D} =  -\frac{3 \hbar \omega_{\perp}^{2} \ln(R_{\perp}/r_{c})}{4 \mu } -  \frac{\hbar k^{2} \ln \left(r_{c} \sqrt{k^{2} - \beta}\right)}{2M} + \Omega.
\end{equation}
This is consistent with previous work \cite{PhysRevLett.84.5919, PhysRevA.63.043608} which shows an off centred vortex in an oblate condensate is straight and has a precession frequency $\Omega -3 \hbar \omega_{\perp}^2\ln(R_{\perp}/r_{c}) / 4 \mu$, in the limit $k\rightarrow 0$.

\begin{figure}[ht]
\centering
\includegraphics[width=0.48\textwidth]{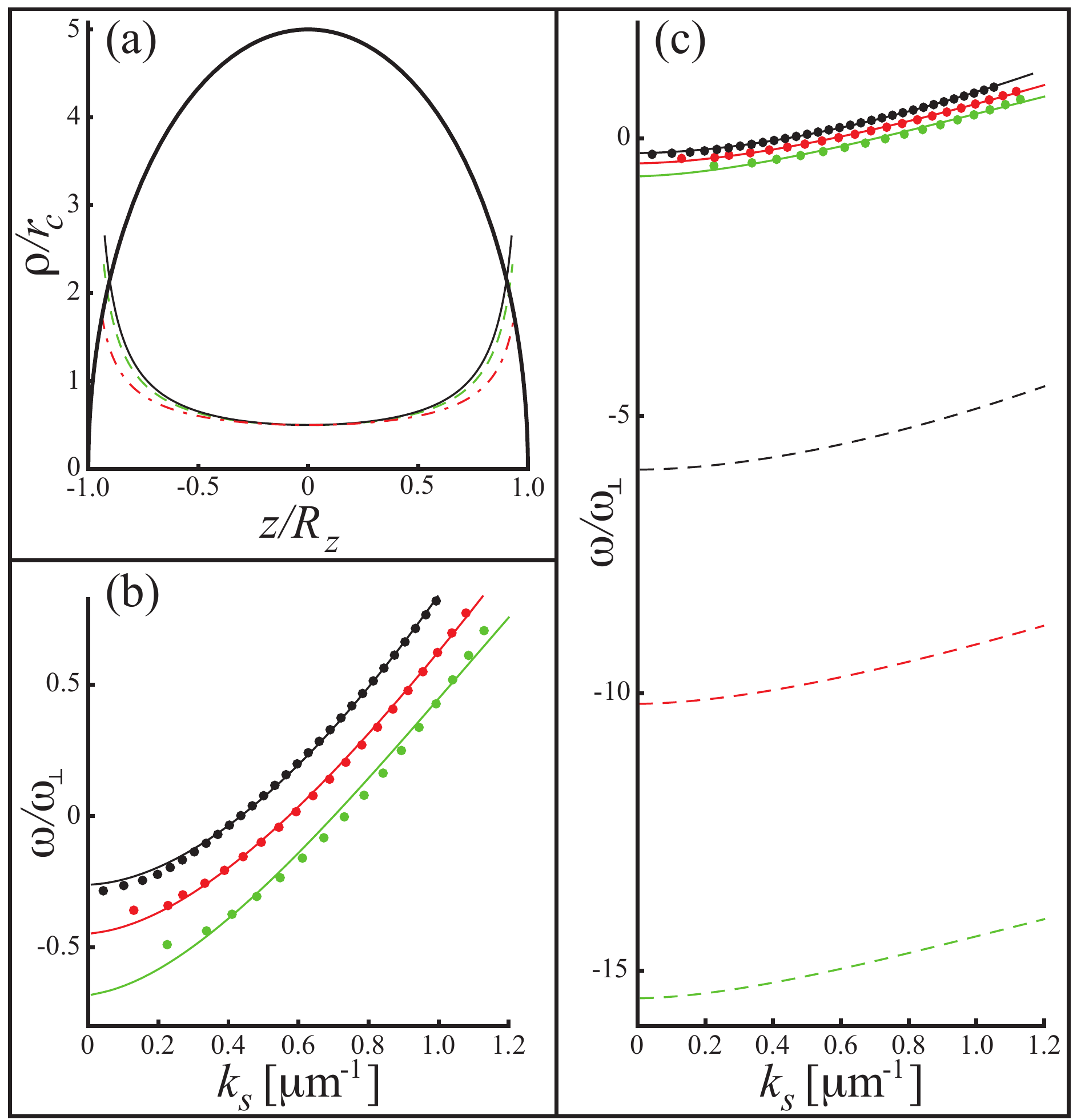}
\caption{(Colour online) (a) The vortex line shape in a prolate BEC, as given by Eq.~\eqref{Andy},  for  $k= 10^{-2}$ $\mu$m$^{-1}$ (red dashed dotted curve), $k=10^{-1}$ $\mu$m$^{-1}$ (green dashed curve) and $k=1$ $\mu$m$^{-1}$ (black solid curve). We have assumed $R_{\perp}/r_{c} =5$, $r_{c} = 0.1$ $\mu$m, $\sqrt{A/\zeta} = \pi/2 $, $C1 =0.5 $, and $C2 =0$.  The thick black solid line defines the Thomas--Fermi surface of the BEC. (b,c) The dispersion relations for Kelvin waves, for $N = 10^{4}$ (lower green realisations), $N= 5 \times 10^{4}$ (middle red realisations), and $N= 5 \times 10^{5}$ (upper back realisations) Rubidium atoms in a harmonic trap: $\omega_{\perp} = 2 \pi \times 98.5$ rad/s and $\omega_{z} = 2 \pi \times 11.8$ rad/s. In (b,c) the numerical calculations (dots) \cite{PhysRevA.78.053604} and analytic results from Eq.~\eqref{disp_Andy} (solid curves) are plotted.
 In (c) the dashed curves show previous analytic results \cite{Phys.Rev.A.62.063617}. For the solid curves and dashed curves in (b,c) the following scaling parameters have been used for $r_c$ and $k$:  $\alpha =$ 0.134, 0.257, and 0.264 for $N = 5 \times10^{5}$, $5 \times 10^{4}$, and $10^{4}$, respectively, and $s = (\lambda R_{z}/ l_{\omega_z})^{1/2}$ for all $N$.}

\label{fig}
\vspace{-0.4cm}
\end{figure}

{\it Prolate Limit --} In the prolate limit ($\lambda \rightarrow 0$), $\gamma$, $\zeta$ and $\varepsilon$ all tend to $\infty$. Hence Eq.~\eqref{s1eq} becomes
\begin{equation}
\rho'^{2} =  \frac{\zeta}{\gamma} \rho' \frac{\partial^{2} \rho'}{\partial z'^{2}} - \frac{\varepsilon}{\gamma} \left(\frac{\partial \rho'}{\partial z'}\right)^{2},
\end{equation}
with the solution
\begin{equation}
\rho'(z') = C_1 \left\{\cos\left[ (z' - \zeta \mbox{ } C_2) \sqrt{\frac{A}{\zeta}}  \right] \right\}^{-\frac{\zeta}{1+\zeta}}. \label{Andy}
\end{equation}
Again $\rho'$ only has a symmetric solution ($C_2=0$). This causes $\rho'$ to have a $1/\cos(z)$ like structure similar to that predicted in simulations \cite{PhysRevA.64.043611,PhysRevA.64.053611}. Typical solutions of Eq.~\eqref{Andy} are plotted in Fig.~\ref{fig} (a), demonstrating the ``U" shape of the radial coordinate. Furthermore as $-\frac{\zeta}{1+\zeta}$ depends on $k$, this ``U" shape will straighten as the momentum of the wave increases, as seen in Fig.~\ref{fig}(a). 

To obtain the dispersion relation for the prolate case  Eq.~\eqref{Andy} needs to be single-valued within the condensate. This gives the condition $\sqrt{A/\zeta} = \pi/2$ or
\begin{gather} 
\omega =  \omega_{2D}  - \frac{\hbar \omega_{z}^{2} \pi^{2} \zeta^{2} }{8 \mu \left(1 + \zeta\right)}, \label{drp}
\end{gather}
which is an approximation to the dispersion relation of helical waves in a prolate trap, and is distinct from the solution for $\omega_z\equiv 0$. This occurs because any trapping in $z$ provides a scale over which the vortex line can curve. Hence, when there is no trapping in $z$, $\varepsilon/\gamma=\zeta/\gamma=0$, turning Eq.~\eqref{s1eq} into $ \gamma \rho'^{2}=0$. Therefore, for the case of  $\omega_z= 0$ the dispersion relation is the same as the oblate limit, given by Eq.~\eqref{dro}, i.e. the vortex line is straight 
\cite{PhysRevLett.84.5919, PhysRevA.63.043608,PhysRevA.86.013605}.



{\it General Solution --} To determine the general dispersion relation from Eq.~\eqref{gen} a similar treatment to the extremely prolate trap is employed. 
As the $\rho'$ solution is a hyper-geometric function to a negative power, the zeros of the hyper-geometric function must lie outside the condensate. 
The zeros of the confluent hyper-geometric function $_{1}F_{1}(l;m;x)$ can be approximated by \cite{NISTHMF}
\begin{equation} \label{zeros}
X_{0} \approx \frac{ \pi^{2} \left( r +\frac{m}{2} - \frac{3}{4}\right)^{2}}{2 m - 4 l},
\end{equation}
where $X_{0}$ is the approximate $x$ value of the $r$th zero. Combining Eq.~\eqref{zeros} with the form of the hyper-geometric function in Eq.~\eqref{gen} and stipulating that the first zero is at $z'=1$ the dispersion relation becomes
\begin{align} 
\omega =  \mbox{ } \omega_{2D} + \frac{ \hbar \omega_{z}^{2} \zeta \left( 2 -  \pi^{2} \zeta \right)}{8 \mu \left(1 + \zeta \right)}. \label{disp_Andy}
\end{align} 
or equivalently, for $\beta /k^2 \rightarrow 0$,
\begin{equation} \label{disp}
\omega =  \omega_{0} +\omega_{1} +\frac{\hbar k^{2} \ln \left( \frac{1}{r_{c} |k|}\right)}{2 M} + \Omega ,
\end{equation}
where
\begin{align}
\omega_{0} =& - \frac{3\hbar  \omega_{\perp}^{2} \ln\left( \frac{R_{\perp}}{ r_{c}} \right)  }{4  \mu ,} \nonumber \\
\omega_{1} =& - \frac{ \hbar \omega_{z}^{2} \ln \left( r_{c} |k|\right)\left[ 4 + \pi^{2} \ln \left( r_{c} |k|\right)\right]}{16 \mu \left[2-\ln \left( r_{c}|k|\right)\right]}. \nonumber
\end{align}
Equations \eqref{disp_Andy} and  \eqref{disp} indicate that helical waves, in a parabolic trap, obey the usual dispersion relation [Eq.~\eqref{hd}], with two constants $\omega_{0}$, from confinement in $\rho$, and $\omega_{1}$, from confinement in $z$. The form of $\omega_{0}$ matches that for extremely oblate traps [Eq.~\eqref{dro}], while $\omega_{1}$ contains new behaviour. Essentially, $\omega_{1}$ is constant, with weak logarithmic dependence on  $k$, and becomes larger as $\omega_z$ increases. As $\lambda \rightarrow \infty$ Eq.~\eqref{disp_Andy} does not replicate the dispersion in the oblate trapping limit. This is because in the very oblate limit Eq.~\eqref{zeros} fails to predict the location of the zeros accurately.

To test the validity of the above analysis we now compare the solutions of Eq.~\eqref{disp_Andy} with numerical calculations \cite{PhysRevA.78.053604}. The dispersion relation of Kelvin waves in a prolate condensate for different particle numbers corresponding to numerical calculations \cite{PhysRevA.78.053604} are shown in Fig.~\ref{fig}(b,c) [dots]. 
Previous analytic predictions in this limit, see Eq.~(70) in Ref.~\cite{Phys.Rev.A.62.063617}, poorly replicated theses results [dashed curves in Fig.~\ref{fig}(c)].
To compare these numerical results with Eq.~\eqref{disp_Andy}, $r_{c}$ needs to first be considered. 

The core parameter, $r_{c}$, implicit in Eq.~\eqref{disp_Andy}, characterizes the vortex core size. In a trapped BEC the healing length, $\xi$, is of order the vortex core radius, which is a function of position. As such we define the core parameter $r_c=\alpha \overline{\xi}$  to be some fraction, $\alpha$, of the healing length averaged over the Thomas--Fermi volume: $\overline{\xi} =(R_{\perp}^{2} R_{z} / 6 a N )^{1/2}$. To compare with numerical results we allow $\alpha$ to be a free parameter, of order $1$. 




In Figs. 1(b,c) (solid curves) we plot the Kelvin wave dispersion,  Eq.~\eqref{disp_Andy}, where we have rescaled $k \rightarrow sk=k_s$. Due the inhomogeneity of the condensate we expect $k$ to vary spatially. As such this rescaling is motivated by the observation \cite{PhysRevA.78.053604} that in numerical calculations $k$ varies along the vortex line.  Interestingly, we find that the matching between analytical results and numerical calculations is optimized for $s = (\lambda R_{z}/ l_{\omega_z})^{1/2}$, where $l_{\omega_z}=\sqrt{\hbar/(m \omega_z)}$ is the harmonic oscillator length in $z$. 

 Figure.~\ref{fig}(c) shows that Eq.~\eqref{disp_Andy} (solid curves) is a vast improvement compared with the previous analytic predictions (dashed curves). Additionally, Fig.~\ref{fig}(b) shows excellent quantitative agreement between the numerical and analytical results, with the agreement improving as the number of particles increases and the BEC becomes more Thomas--Fermi like. 
The general features of the dispersion relation are dominated by: (i) a frequency shift and (ii) a functional form $\propto -k^2 \ln r_c k$. The shift is dominated by the confinement in $\rho$, given by $\omega_0$, with a small contribution from the confinement in $z$, given by the second term in Eq.~\eqref{disp_Andy}. The functional dependence of $\omega$ is essentially dominated by the solution for a vortex line in an un-trapped condensate, Eq.~\eqref{hd}, also with a small correction arising from the confinement in $z$, given by the second term in Eq.~\eqref{disp_Andy}.  We note that the self-consistent determination of $\beta$ only influences the dispersion for $k<10^{-4}$ $\mu$m$^{-1}$ and hence, in Figs. ~\ref{fig}(b,c), plays no role over the wave-vectors considered. 

In summary, we have analytically determined the Kelvin wave dispersion relation for a vortex line in a BEC trapped in a cylindrically symmetric parabolic trap, Eq.~\eqref{disp_Andy}. This result quantitatively agrees with numerical calculations, in stark contrast to  previous descriptions. We also find that the dispersion relation in the oblate trapping limit, Eq.~\eqref{dro}, coincides with previous analytical work. In general these results are derived from the governing equation of motion for Kelvin waves on a quantized vortex line, Eq.~\eqref{s1eq}. Equation~\eqref{s1eq} successfully predicts the behaviour in very oblate and prolate condensates, and mathematically shows the link between the vortex line shape with no trapping in $z$ and strong trapping in $z$.
Within the context of  experimental activity in the study of excitations of quantized vortex lines in trapped BECs \cite{PhysRevLett.85.2223,PhysRevLett.90.100403,PhysRevLett.86.2922,PhysRevLett.89.200403},  this work provides a simple analytic tool for the analysis of Kelvin waves. Additionally, given the close agreement between numerical calculations and Eq.~\eqref{disp_Andy} we expect future experimental measurements of the Kelvin wave spectra  to be in close agreement with our analytical description.



\bibliography{Bibliography}

\begin{thebibliography}{37}
\bibitem[{Not()}]{NoteLyndon}
\bibinfo{note}{From Eqs. (68) and (69) in Ref. [28] it is possible to determine
  some of the dependence of the unknown constant to obtain Eqs. (2) and (3).
  This process is illustrated in Sec. VII of Ref. [28].}
\end{thebibliography}
\end{document}